\documentclass[%
 reprint,
 aps,
 superscriptaddress,
 raggedbottom,
 prl]{revtex4-2}
\usepackage{CJK}

\usepackage{amsmath,amssymb,mathtools,extarrows,mathrsfs}
\usepackage{tikz,float,subcaption,graphbox}

\usepackage{titlesec}
\titleformat{\section}{\centering\normalsize\normalfont\bf}{\thesection}{0em}{}
\usepackage{hyperref}

\begin{document}
\begin{CJK*}{UTF8}{gbsn}
\title{A Hidden Permutation Symmetry of Squared Amplitudes in ABJM Theory}

\author{Song He (何颂)}
\email{songhe@itp.ac.cn}
\affiliation{Institute of Theoretical Physics, Chinese Academy of Sciences, Beijing 100190, China}
\affiliation{School of Fundamental Physics and Mathematical Sciences, Hangzhou Institute for Advanced Study, Hangzhou, Zhejiang 310024, China}
\author{Canxin Shi (施灿欣)}
\email{shicanxin@itp.ac.cn}
\affiliation{Institute of Theoretical Physics, Chinese Academy of Sciences, Beijing 100190, China}
\author{Yichao Tang (唐一朝)}
\email{tangyichao@itp.ac.cn}
\affiliation{Institute of Theoretical Physics, Chinese Academy of Sciences, Beijing 100190, China}
\affiliation{School of Physical Sciences, University of Chinese Academy of Sciences, Beijing 100049, China}
\author{Yao-Qi Zhang (张耀奇)}
\email{zhangyaoqi@itp.ac.cn}
\affiliation{Institute of Theoretical Physics, Chinese Academy of Sciences, Beijing 100190, China}
\affiliation{School of Physical Sciences, University of Chinese Academy of Sciences, Beijing 100049, China}
\date{\today}

\begin{abstract}
We define the \emph{square amplitudes} in planar Aharony-Bergman-Jafferis-Maldacena theory (ABJM), analogous to that in ${\cal N}{=}4$ super-Yang-Mills theory (SYM). Surprisingly, the $n$-point $L$-loop integrands with fixed $N{:=}n{+}L$ are unified in a single generating function. Similar to the SYM four-point half-BPS correlator integrand, the generating function enjoys a hidden $S_N$ permutation symmetry in the dual space, allowing us to write it as a linear combination of weight-3 planar $f$-graphs. Remarkably, through Gram identities it can also be represented as a linear combination of \emph{bipartite} $f$-graphs which manifest the important property that no odd-multiplicity amplitude exists in the theory. The generating function and these properties are explicitly checked against squared amplitudes for all $n$ with $N{=}4,6,8$. By drawing analogies with SYM, we conjecture some \emph{graphical rules} the generating function satisfy, and exploit them to bootstrap a unique $N{=}10$ result, which provides new results for $n{=}10$ squared tree amplitudes, as well as integrands for $(n,L){=}(4,6),(6,4)$. Our results strongly suggest the existence of a ``bipartite correlator'' in ABJM theory that unifies all squared amplitudes and satisfies physical constraints underlying these graphical rules. 
\end{abstract}

\maketitle
\end{CJK*}

\section{Introduction}

Planar $\mathcal N{=}4$ super-Yang-Mills theory (SYM) with gauge group $SU(N_c)$ is a rare example of a four-dimensional integrable theory, enabling the nonperturbative study of conformal data and many off-shell observables~\cite{Beisert:2010jr,Basso:2015zoa,Coronado:2018ypq,Belitsky:2020qrm,Belitsky:2020qir}. On the other hand, as the supersymmetric cousin of the phenomenogically interesting QCD, scattering amplitudes~\cite{Travaglini:2022uwo} in SYM represent another class of important observables. For amplitudes, integrability is reflected in the dual superconformal and Yangian symmetry~\cite{Drummond:2008vq,Brandhuber:2008pf,Drummond:2008bq,Drummond:2009fd,Bargheer:2009qu}. Although infrared divergences~\cite{Brandhuber:2009kh,Beisert:2010gn,Caron-Huot:2011dec,Bullimore:2011kg} obscure dual conformal invariance (DCI), the loop \emph{integrand} (well-defined in the planar limit) manifests the Yangian symmetry~\cite{Arkani-Hamed:2012zlh}. Physically, dual superconformal symmetry of amplitudes is understood as ordinary superconformal symmetry of null polygonal Wilson loops through the amplitude/Wilson loop/correlator triality~\cite{Heslop:2022xgp}.

In particular, it is possible to introduce the \emph{squared amplitude}, a bosonic object simpler than the (super)amplitude itself, which is believed to encode all information of the amplitude~\cite{Ambrosio:2013pba,Heslop:2018zut}. Moreover, the squared amplitude and the analogous squared form factor~\cite{He:2025zbz} are key ingredients in perturbative calculations such as those for energy correlators~\cite{Moult:2025nhu}. At the integrand level, the triality relates the squared amplitude to the adjoint Wilson loop and the lightlike limit of the four-point half-BPS correlator. The method of Lagrangian insertion~\cite{Eden:2011yp} reveals a hidden permutation symmetry~\cite{Eden:2011we} of the correlator integrand, which enables the efficient bootstrap~\cite{Bourjaily:2015bpz,Bourjaily:2016evz,He:2024cej,Bourjaily:2025iad} of the correlator integrand (and hence the squared amplitude) using the so-called $f$-graphs.

The Aharony-Bergman-Jafferis-Maldacena theory (ABJM)~\cite{Aharony:2008ug} is a three-dimensional $\mathcal N{=}6$ superconformal Chern-Simons-matter theory, originally introduced to study the dynamics of M2-branes. It was soon realized that the theory is very similar to SYM in the sense that it is also integrable in the planar limit~\cite{Klose:2010ki}. In particular, the (integrand of) scattering amplitudes also enjoy dual superconformal and Yangian symmetry~\cite{Bargheer:2010hn,Huang:2010qy,Gang:2010gy,Chen:2011vv,Lee:2010du,Huang:2014xza,Bianchi:2014iia}. However, the physical origin of DCI is not as clear as in SYM. Although there are some hints of duality to the Wilson loop at 4 points~\cite{Henn:2010ps,Bianchi:2011rn,Bianchi:2011dg}, attempting to identify a dual ``super Wilson loop'' (necessary at higher points due to a lack of bosonic MHV sectors in the theory) is met with difficulties~\cite{Bargheer:2012cp,Berkovits:2008ic,Colgain:2016gdj,Rosso:2014oha}. Also, explicit results of ABJM amplitudes are not as satisfying as in SYM. Using generalized unitarity~\cite{Britto:2004nc} and geometry~\cite{Arkani-Hamed:2013jha,Arkani-Hamed:2021iya}, it is possible to obtain the 2-loop 8-point integrand~\cite{Bianchi:2012cq,Brandhuber:2012un,Caron-Huot:2012sos,He:2022lfz,He:2023rou} and the 5-loop 4-point integrand~\cite{He:2022cup}.

In this Letter, we take a major step towards completing the parallel story between ABJM and SYM by providing evidence of a duality between the squared amplitude and some ``correlator''. Specifically, by carefully defining the (integrand of) squared amplitude, we observe that it can be written as the lightlike limit of a \emph{generating function} which also enjoys hidden permutation symmetry. Moreover, the $f$-graph representation is manifestly \emph{bipartite}, echoing previous studies of the (logarithm of) amplitude~\cite{He:2022cup,He:2023rou}. This strongly suggests the interpretation of the generating function as some ``bipartite correlator with Lagrangian insertions''. Based on the physical picture and drawing analogies from SYM, we conjecture some \emph{graphical rules} the generating function should satisfy. Together with the bipartite property, this enables us to bootstrap the generating function at 10 points, obtaining new results for tree-level 10-point, 2-loop 8-point, and 4-loop 6-point squared amplitudes, as well as the 6-loop 4-point (un-squared) amplitude. 

\section{Squared amplitudes in SYM}

Let us briefly review the SYM story. In chiral on-shell superspace~\cite{Arkani-Hamed:2012zlh,Elvang:2015rqa}, Poincar\`e supersymmetry implies that the superamplitude $\mathcal A_n(\lambda,\tilde\lambda,\tilde\eta){=}\delta^4(P)\delta^8(Q)A_n(\lambda,\tilde\lambda,\tilde\eta)$ satisfies $\widetilde QA_n{=}0$, where ($\partial_{iA}{:=}\partial_{\tilde\eta_i^A}$)
\begin{equation}
    P^{\alpha\dot\alpha}{=}\sum_{i=1}^n\lambda_i^\alpha\tilde\lambda_i^{\dot\alpha},\ Q^{\alpha A}{=}\sum_{i=1}^n\lambda_i^\alpha\tilde\eta_i^A,\ \widetilde Q^{\dot\alpha}_A{=}\sum_{i=1}^n\tilde\lambda_i^{\dot\alpha}\partial_{iA}.
\end{equation}
Parity maps the superamplitude $\mathcal A_n(\lambda,\tilde\lambda,\tilde\eta)$ to a differential operator $\overline{\mathcal A}_n(\tilde\lambda,\lambda,\partial_{\tilde\eta}){=}\delta^4(P)\delta^8(\widetilde Q)\overline A_n(\tilde\lambda,\lambda,\partial_{\tilde\eta})$ on chiral superspace. The squared amplitude is then defined on the support of $\delta(P)$ as
\begin{equation}\label{eq:symSquare}
    M_n=\frac12\overline A_n(\tilde\lambda,\lambda,\partial_{\tilde\eta})A_n(\lambda,\tilde\lambda,\tilde\eta)\Big|_{\tilde\eta,\partial_{\tilde\eta}=0}.
\end{equation}
This can be computed by writing $A_n{=}\sum f(\lambda,\tilde\lambda)\prod\tilde\eta$ as a polynomial in $\tilde\eta$. In $\overline A_nA_n$, all cross terms vanish upon setting $\tilde\eta,\partial_{\tilde\eta}{=}0$, and $M_n{=}\frac12\sum f(\tilde\lambda,\lambda)f(\lambda,\tilde\lambda)$ is simply the sum of the squared coefficients, which is parity-even and hence a function of Mandelstam variables only.

Note that it is necessary to strip off $\delta(P)\delta(Q)$ in~\eqref{eq:symSquare}, because $\overline{\mathcal A}_n\mathcal A_n{\propto}\widetilde Q\mathcal A_n{=}0$ trivially. Moreover, \eqref{eq:symSquare} is well-defined in the sense that $M_n$ is invariant under shifting $A_n{\mapsto}A_n'{=}A_n{+}QX$ keeping $\widetilde QA_n'{=}0$, because the would-be cross terms vanish due to $\widetilde QA_n'{=}0$.

In the planar limit $N_c{\to}\infty$ with fixed 't Hooft coupling $a{=}g_{\rm YM}^2N_c$, it is possible to define a loop integrand $A_n^{(L)}(\lambda,\tilde\lambda,\tilde\eta;y_1,\cdots,y_L)$ where the loop momenta are specified by the dual variables $y_\ell\in\mathbb R^4$:
\begin{equation}
    A_n=\sum_{L=0}^\infty a^L\int{\rm d}^{4L}y\,A_n^{(L)},
\end{equation}
where $A_n^{(L)}$ is defined to be symmetric under permuting $y_1,{\cdots},y_L$ because the integration region is symmetric. The integrand $M_n^{(L)}$ of squared amplitudes is given by
\begin{align}
    M_n^{(L)}=\frac12&\sum_{\ell=0}^L\frac1{L!}\sum_{\sigma\in S_L}A_n^{(\ell)}(\tilde\lambda,\lambda,\partial_{\tilde\eta};y_{\sigma(1)},\cdots,y_{\sigma(\ell)})\nonumber\\
    \times&A_n^{(L-\ell)}(\lambda,\tilde\lambda,\tilde\eta;y_{\sigma(\ell+1)},\cdots,y_{(L)})\Big|_{\tilde\eta,\partial_{\tilde\eta}=0}.
\end{align}
Introducing dual points $x_i{\in}\mathbb R^4$ such that $x_i{-}x_{i+1}{=}\lambda_i\tilde\lambda_i$, we can rewrite $M_n^{(L)}(x_i,y_\ell)$ as a rational function of the dual points. Remarkably, there exists a single object $F_N(x_1,{\cdots},x_N)$ with $N{:=}n{+}L$ that packages $M_n^{(L)}$ with different $n,L$ through $n$-gon lightlike limits~\cite{Bourjaily:2016evz}:
\begin{equation}\label{eq:symFM}
    \lim_{x_{12}^2,\cdots,x_{n1}^2\to0}\sigma_nF_N=M_n^{(L)}(x_1,\cdots,x_N),
\end{equation}
where $\sigma_n{:=}x_{12}^2x_{23}^2{\cdots}x_{n1}^2$. Moreover, $F_N$ is invariant under permutation of $x_1,{\cdots},x_N$, unifying ``loops and legs'' and allowing the introduction of weight-4 $f$-graphs~\cite{Eden:2011we}.

We remark that in the literature, the SYM squared amplitude is usually defined~\cite{Bourjaily:2016evz,Heslop:2022xgp} in terms of the stripped N$^k$MHV amplitudes $R_{n,k}^{(L)}{:=}\langle12\rangle\cdots\langle n1\rangle A_n^{(L)}|_{\tilde\eta^{4k}}$, which agrees with~\eqref{eq:symSquare} up to a Parke-Taylor factor:
\begin{equation}\label{eq:differentSqDef}
    \widetilde M_n^{(L)}=\frac12\sum_{k=0}^{n-4}\sum_{\ell=0}^L\frac{R_{n,k}^{(\ell)}R_{n,n-4-k}^{(L-\ell)}}{R_{n,n-4}^{(0)}}=M_n^{(L)}\prod_{i=1}^ns_{i,i+1}.
\end{equation}

\section{Squared amplitudes in ABJM}

In ABJM, the (integrand of) squared amplitudes can be defined in a similar fashion. Under the gauge group $U(N_c)_{k_{\rm CS}}{\times}U(N_c)_{-k_{\rm CS}}$, the two on-shell superfields~\cite{Elvang:2015rqa} transform as $\Phi{\in}(\mathbf N_c,\overline{\mathbf N}_c)$ and $\Psi{\in}(\overline{\mathbf N}_c,\mathbf N_c)$. In the planar limit $N_c{\to}\infty$ with fixed 't Hooft coupling $a{=}N_c/k_{\rm CS}$, the color-ordered superamplitude is defined as
\begin{equation}
    \mathcal A_n(\lambda,\eta)=\mathcal A_n(\Psi_1\Phi_2\cdots\Psi_{n-1}\Phi_n),
\end{equation}
where only an even number $n{=}2k$ of particles have nonvanishing amplitudes. Poincar\'e supersymmetry implies that $\mathcal A_n(\lambda,\eta){=}\delta^3(P)\delta^6(Q)A_n(\lambda,\eta)$ satisfies $\widetilde QA_n{=}0$, where $A_n{\sim}\eta^{3(k-2)}$ and ($\partial_{iI}{:=}\partial_{\eta_i^I}$)
\begin{equation}
    P^{\alpha\beta}{=}\sum_{i=1}^n\lambda_i^\alpha\lambda_i^\beta,\ Q^{\alpha I}{=}\sum_{i=1}^n\lambda_i^\alpha\eta_i^I,\ \widetilde Q^\alpha_I{=}\sum_{i=1}^n\lambda_i^\alpha\partial_{iI}.
\end{equation}
Here, $I{=}1,2,3$ manifests the $SU(3)$ subgroup of the R-symmetry $SU(4){\cong}SO(6)$. Parity is trivial in three dimensions and there is only one type of spinor $\lambda$. In order to define the squared amplitude, the correct prescription is charge conjugation~\footnote{See the discussion below eq.(11.51) of~\cite{Elvang:2015rqa}.}, which maps $\mathcal A_n(\lambda,\eta)$ to a differential operator $\overline{\mathcal A}_n(\lambda,\partial_\eta){=}\delta^3(P)\delta^6(\widetilde Q)\overline A_n(\lambda,\partial_\eta)$. As charge conjugation changes $k_{\rm CS}{\leftrightarrow}-k_{\rm CS}$, the integrand $\overline A_n^{(L)}$ is related to $A_n^{(L)}$ through an additional $(-)^L$ factor:
\begin{equation}
    \overline A_n^{(L)}(\lambda,\partial_\eta)=(-)^LA_n^{(L)}(\lambda,\eta)\Big|_{\eta\mapsto\partial_\eta}.
\end{equation}
The squared amplitude is defined as in~\eqref{eq:symSquare}, which at the integrand level reads
\begin{align}\label{eq:defMnL}
    M_n^{(L)}=\frac12&\sum_{\ell=0}^L\frac1{L!}\sum_{\sigma\in S_L}(-)^\ell A_n^{(\ell)}(\lambda,\partial_\eta;y_{\sigma(1)},\cdots,y_{\sigma(\ell)})\nonumber\\
    \times&A_n^{(L-\ell)}(\lambda,\eta;y_{\sigma(\ell+1)},\cdots,y_{(L)})\Big|_{\eta,\partial_\eta=0}.
\end{align}
Due to the $(-)^\ell$ factor, $M_n^{(L)}$ is only nonzero if $L$ is even. We will see that an $S_N$-permutation invariant generating function $F_N(x_1,{\cdots},x_N)$ with $N{:=}n{+}L$ unifies $M_n^{(L)}$ in the same way~\eqref{eq:symFM} as in SYM. As the simplest case, the $4$-point tree amplitude $A_4^{(0)}{=}\frac1{\langle12\rangle\langle23\rangle}$ leads to
\begin{equation}
    M_4^{(0)}=\frac1{2x_{13}^2x_{24}^2}=\lim_{x_{12}^2,\cdots,x_{41}^2\to0}\sigma_4\underbrace{\left(\frac12\frac1{x_{12}^2x_{13}^2x_{14}^2x_{23}^2x_{24}^2x_{14}^2}\right)}_{F_4}.
\end{equation}

\section{$N=6$: a first indication}

Let us proceed to the $N{=}6$ case. For $(n,L){=}(4,2)$, the amplitude $A_4$ is $\eta$-independent. Hence,~\eqref{eq:defMnL} reads
\begin{equation}\label{eq:M42def}
    M_4^{(2)}{=}A_4^{(0)}()\frac{A_4^{(2)}(56){+}A_4^{(2)}(65)}{2}{-}\frac12A_4^{(1)}(5)A_4^{(1)}(6),
\end{equation}
where $A_4^{(L)}(5,{\cdots},4{+}L)$ denotes the integrand with dual loop momenta $x_5,{\cdots},x_{4+L}$, whose explicit expressions can be found in~\cite{Chen:2011vv}. Explicit computation gives~\footnote{It is extremely difficult to align the conventions of different papers. Therefore, we ignore the overall normalization of $M_n^{(L)}$ in this Letter. However, we believe there exists a good convention such that~\eqref{eq:symFM} holds exactly in ABJM.}
\begin{align}\label{eq:M42}
    M_4^{(2)}&=\left[\frac1{x_{15}^2x_{25}^2x_{56}^2x_{36}^2x_{46}^2}+\text{cyc}(1234)\right]\nonumber\\
    &-{\color{blue}\left[\frac{x_{13}^2}{x_{15}^2x_{16}^2x_{25}^2x_{35}^2x_{36}^2x_{46}^2}+(13\leftrightarrow24)+(5\leftrightarrow6)\right]}\nonumber\\
    &-{\color{red}\left[\frac{x_{13}^2/x_{24}^2}{x_{15}^2x_{16}^2x_{35}^2x_{36}^2x_{56}^2}+(13\leftrightarrow24)\right]}\nonumber\\
    &+\frac12\left[\frac{x_{13}^2x_{26}^2x_{45}^2/x_{24}^2}{x_{15}^2x_{25}^2x_{35}^2x_{56}^2x_{36}^2x_{46}^2x_{16}^2}+\text{cyc}(1234)\right]\nonumber\\
    &+\frac12\frac{x_{13}^2x_{24}^2x_{56}^2}{x_{15}^2x_{25}^2x_{35}^2x_{45}^2x_{16}^2x_{26}^2x_{36}^2x_{46}^2}.
\end{align}
By dividing $\sigma_4$, one can easily obtain $F_6$ as a sum of weight-3 $f$-graphs, whose leading term in the 4-gon lightlike limit $x_{12}^2,{\cdots},x_{41}^2{\to}0$ coincides with $M_4^{(2)}$:
\begin{equation}\label{eq:F6p}
    F_6=\includegraphics[align=c,scale=0.25]{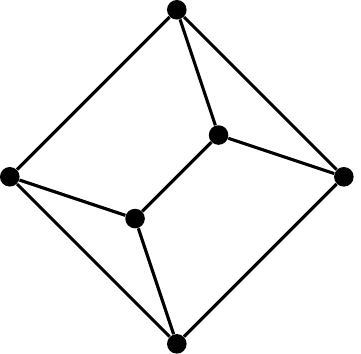}-\includegraphics[align=c,scale=0.25]{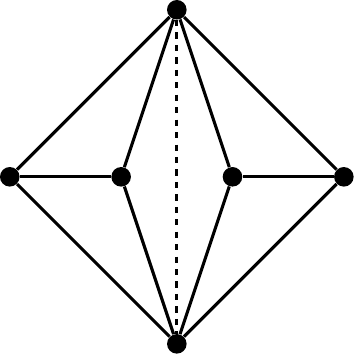}+\frac12\includegraphics[align=c,scale=0.25]{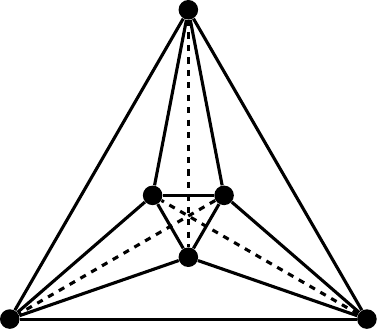}.
\end{equation}
Here, each $f$-graph denotes a sum of permutation inequivalent monomials where each solid line denotes a denominator $1/x_{ij}^2$ and each dashed line denotes a numerator $x_{ij}^2$. The conformal weight of each vertex (the number of solid edges minus the number of dashed lines) is always $3$, as expected from DCI in three dimensions. Each $f$-graph is $S_N$-permutation invariant. For example,
\begin{equation}
    \includegraphics[align=c,scale=0.25]{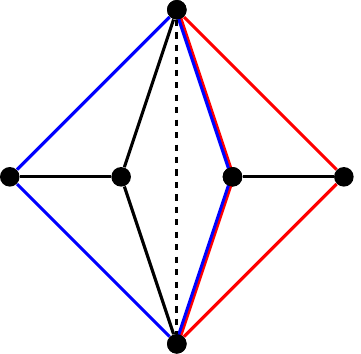}=\underbrace{\frac{x_{56}^2}{x_{15}^2x_{25}^2x_{35}^2x_{45}^2x_{16}^2x_{26}^2x_{36}^2x_{46}^2x_{12}^2x_{34}^2}+\text{perm}}_{\substack{\text{$45=6!/16$ terms in total,}\\\text{where $16$ is the symmetry factor of the graph}}}.
\end{equation}
Monomials surviving the 4-gon lightlike limit correspond to solid line 4-cycles; above, we have shown the two inequivalent 4-cycles and their contributions in~\eqref{eq:M42} with the same colors. Note that all $f$-graphs in~\eqref{eq:F6p} are planar in the sense that the solid line subgraphs are planar. This is not surprising, as the integrands in~\cite{Chen:2011vv} are planar Feynman diagrams.

For $(n,L){=}(6,0)$, the amplitude $A_6{\sim}\eta^3$~\cite{Gang:2010gy}:
\begin{equation}
    A_6^{(0)}=f_+(\lambda)\delta^3\left(\sum_{i=1}^6\alpha_+^i(\lambda)\eta_i\right)+f_-(\lambda)\delta^3\left(\sum_{i=1}^6\alpha_-^i(\lambda)\eta_i\right).
\end{equation}
According to the definition,
\begin{align}\label{eq:M60}
    M_6^{(0)}&=\frac12\sum_{i_1i_2i_3=1}^6\left(f_+\alpha_+^{i_1}\alpha_+^{i_2}\alpha_+^{i_3}+f_-\alpha_-^{i_1}\alpha_-^{i_2}\alpha_-^{i_3}\right)^2\nonumber\\
    &=\frac2{x_{14}^2x_{25}^2x_{36}^2}=\lim_{x_{12}^2,\cdots,x_{61}^2\to0}\sigma_6\,2\includegraphics[align=c,scale=0.25]{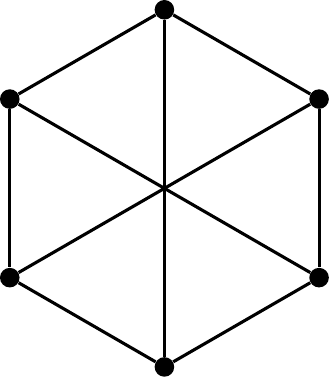}.
\end{align}
This $f$-graph is not planar; however, it is bipartite (also referring to the solid line subgraph). It looks different from~\eqref{eq:F6p}, but in three dimensions, due to the vanishing of the conformal Gram determinant $\det[x_{ij}^2]_{i,j=1}^6{=}0$~\cite{Eden:2012tu},
\begin{equation}
    2\includegraphics[align=c,scale=0.25]{images/bf6_1.pdf}=\includegraphics[align=c,scale=0.25]{images/pf6_1.pdf}-\includegraphics[align=c,scale=0.25]{images/pf6_2.pdf}+\frac12\includegraphics[align=c,scale=0.25]{images/pf6_3.pdf}.
\end{equation}
This demonstrates that both $M_4^{(2)}$ and $M_6^{(0)}$ are unified in the same generating function $F_6$, which can be represented as a linear combination of either planar $f$-graphs or bipartite $f$-graphs, related through three-dimensional Gram identities. Incidentally, directly using the integrands for $\log(A_4/A_4^{(0)})$~\cite{He:2022cup}, we would obtain~\footnote{From $M_4{=}\frac12[A_4^{(0)}]^2\exp\log(A_4(-a)A_4(a)/[A_4^{(0)}]^2)$, we can express $M_4^{(L)}$ in terms of $\underline{\tilde\Omega}_\ell$ with even $\ell$ only, where $\tilde\Omega_\ell/\ell!$ is the $\ell$-loop integrand of $\log(A_4(a)/A_4^{(0)})$~\cite{He:2022cup}.} $M_4^{(2)}{=}\frac2{x_{15}^2x_{35}^2x_{56}^2x_{26}^2x_{46}^2}{+}(5\leftrightarrow6){=}\lim\sigma_4\,2\,\includegraphics[align=c,scale=0.07]{images/bf6_1.pdf}$ which manifests the bipartite pole structure. Note that it is natural to consider bipartite $f$-graphs instead of just planar $f$-graphs in ABJM, since these contain no odd-length cycles and manifest the vanishing of odd-$n$ (squared) amplitudes through $\lim\sigma_nF_N=0$~\footnote{The lowest-point $F_4{=}\frac12\,\includegraphics[align=c,scale=0.07]{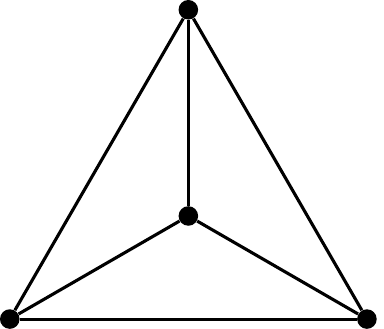}$ is an exception by being planar but not bipartite. This is similar to SYM where $F_{N\geq6}^{\rm SYM}$ are planar but $F_5^{\rm SYM}$ is nonplanar.}.

\section{$N=8$: bipartite vs. planar}

The lesson learned at $N{=}6$ applies to $N{=}8$. For example, the $6$-point $L$-loop amplitudes can be written as~\cite{Caron-Huot:2012sos}:
\begin{equation}
    A_6^{(L)}= I_6^{(L)}A_6^{(0)}+J_6^{(L)}A_{6,{\rm shift}}^{(0)}.
\end{equation}
Here, $I_6^{(L)}$ and $J_6^{(L)}$ are purely bosonic functions, and all the $\eta$-dependence are captured by the tree amplitude $A_6^{(0)}$ and its cyclic-by-one image:
\begin{equation}
    A_{6,{\rm shift}}^{(0)}(\Psi_1\Phi_2\cdots\Psi_5\Phi_6):=A_6^{(0)}(\Psi_2\Phi_3\cdots\Psi_6\Phi_1).
\end{equation}
Computation shows that
\begin{align}
    \overline A_6^{(0)}(\partial_\eta)A_6^{(0)}(\eta)=\overline A_{6,{\rm shift}}^{(0)}(\partial_\eta)A_{6,{\rm shift}}^{(0)}(\eta)&=2M_6^{(0)},\\
    \overline A_6^{(0)}(\partial_\eta)A_{6,{\rm shift}}^{(0)}(\eta)=\overline A_{6,{\rm shift}}^{(0)}(\partial_\eta)A_6^{(0)}(\eta)&=0.
\end{align}
Therefore, cross terms $I\times J$ do not contribute to $M_6^{(2)}$:
\begin{equation}
    \frac{M_6^{(2)}}{M_6^{(0)}}=I_6^{(2)}(78) +I_6^{(2)}(87)-I_6^{(1)}(7)I_6^{(1)}(8)-J_6^{(1)}(7)J_6^{(1)}(8).
\end{equation}

Similarly, we compute $M_4^{(4)}$~\cite{He:2022cup} and $M_8^{(0)}$~\cite{Brandhuber:2012un} and check numerically that $M_4^{(4)},M_6^{(2)},M_8^{(0)}$ are all unified in a generating function $F_8$ through~\eqref{eq:symFM}, where
\begin{equation}\label{eq:F8b}
    F_8=8\includegraphics[align=c,scale=0.25]{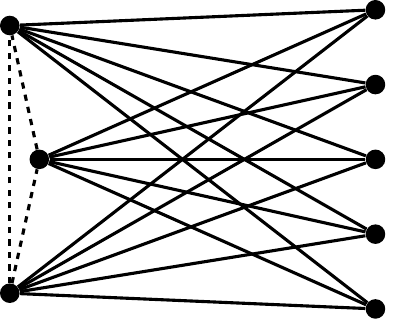}-4\includegraphics[align=c,scale=0.25]{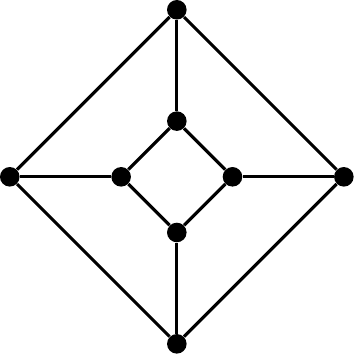}+2\includegraphics[align=c,scale=0.25]{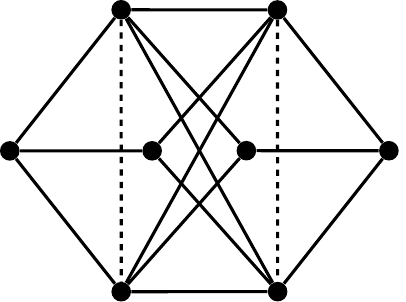}.
\end{equation}
Once again, we see that $F_8$ can be represented graphically as a linear combination of bipartite $f$-graphs.

Let us now discuss the relation between planar $f$-graphs and bipartite $f$-graphs in more detail. It is easy to generate all planar or bipartite $f$-graphs using softwares such as \texttt{plantri}~\cite{plantripaper} and \texttt{nauty}~\cite{MCKAY201494}. At $N{=}8$, there are 61 planar $f$-graphs and 4 bipartite $f$-graphs. Apart from $f_8^{1,2,3}$ shown in~\eqref{eq:F8b}, the fourth bipartite $f$-graph is
\begin{equation}\label{eq:bf8_4}
    f_8^4=\includegraphics[align=c,scale=0.25]{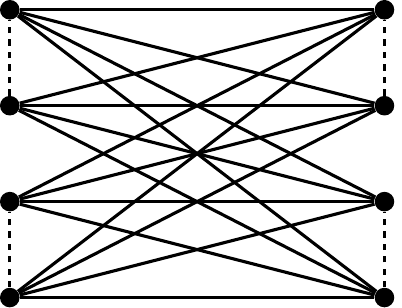}.
\end{equation}
Among the bipartite $f$-graphs, $f_8^2{=}\includegraphics[align=c,scale=0.07]{images/bf8_2.pdf}$ is special in that it is both planar and bipartite. More generally, through Gram identities, certain linear combinations of planar $f$-graphs coincide with linear combinations of bipartite $f$-graphs, as we have seen for $F_6$. In other words, given the vector spaces $\mathcal B,\mathcal P$ generated by bipartite and planar $f$-graphs, we have $F_6{\in}\mathcal B{\cap}\mathcal P$. We check numerically that $F_8{\in}\mathcal B{\cap}\mathcal P$ as well.

With the above observation, we turn the logic around to bootstrap $F_N$, assuming $F_N{\in}\mathcal B{\cap}\mathcal P$. We can numerically identify all linear relations relating planar $f$-graphs to bipartite $f$-graphs, which generate the subspace $\mathcal B{\cap}\mathcal P$. This fixes $F_6{\propto}\includegraphics[align=c,scale=0.07]{images/bf6_1.pdf}$ and constrains $F_8$ to be
\begin{align}\label{eq:F8boots}
    F_8& = c \left(4\includegraphics[align=c,scale=0.25]{images/bf8_1.pdf}+\includegraphics[align=c,scale=0.25]{images/bf8_3.pdf}\right)+c'\includegraphics[align=c,scale=0.25]{images/bf8_2.pdf}\nonumber\\
    &+0\includegraphics[align=c,scale=0.25]{images/bf8_4.pdf}.
\end{align}
In particular, the coefficient of~\eqref{eq:bf8_4} in $F_8$ is fixed to be 0. Note that planarity only implies $F_8{\in}\mathcal P$ which leads to a 61-term ansatz, while the bipartite property drastically reduces the ansatz size down to 2! This demonstrates the impressive constraining power of the $\mathcal B{\cap}\mathcal P$ bootstrap.

Apart from the inter-class Gram identities generating $\mathcal B{\cap}\mathcal P$, there could be intra-class Gram identities within $\mathcal B$ or $\mathcal P$. For example, at $N{=}8$, even though there are 61 planar $f$-graph, actually $\dim\mathcal P{=}60$; the unique linear relation (involving 43 planar $f$-graphs) is recorded in an ancillary file. As a result, although $F_8$ has a unique expansion onto bipartite $f$-graphs~\eqref{eq:F8b}, the expansion onto planar $f$-graphs is not unique.

\section{$N=10$: putative graphical rules}

\definecolor{qcolor}{RGB}{255,127,0}
\definecolor{ecolor}{RGB}{127,0,255}
\newcommand{\dq}{\includegraphics[scale=0.25,align=c]{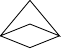}}
\definecolor{dcolor}{RGB}{255,0,127}

Seeing that $F_8{\in}\mathcal B{\cap}\mathcal P$ already fixes $F_8$ down to two coefficients (only one if we ignore the overall normalization), it is natural to ask whether we can fix $F_8$ completely.

Let us recall how $F_N$ is bootstrapped in SYM. There, by studying the asymptotic behavior of the correlator under the cusp~\cite{Alday:2010zy,He:2024cej} (resp., OPE~\cite{Eden:2012tu}) limit, one can extract the double-triangle (resp., triangle) rule that relates $F_N$ and $F_{N-1}$, which we summarize in the Supplemental Material. Notably, it is possible to impose these constraints BEFORE identifying linear combinations of $f$-graphs related by Gram identities, which makes them \emph{graphical rules}.

\begin{figure}[H]
    \centering
    \subcaptionbox{Quadrangle rule:\\$\alpha(\frac{c_2}8+\frac{c_3}2+\frac{c_4}8)=2\frac{b_1}8$}{\includegraphics[width=0.48\linewidth]{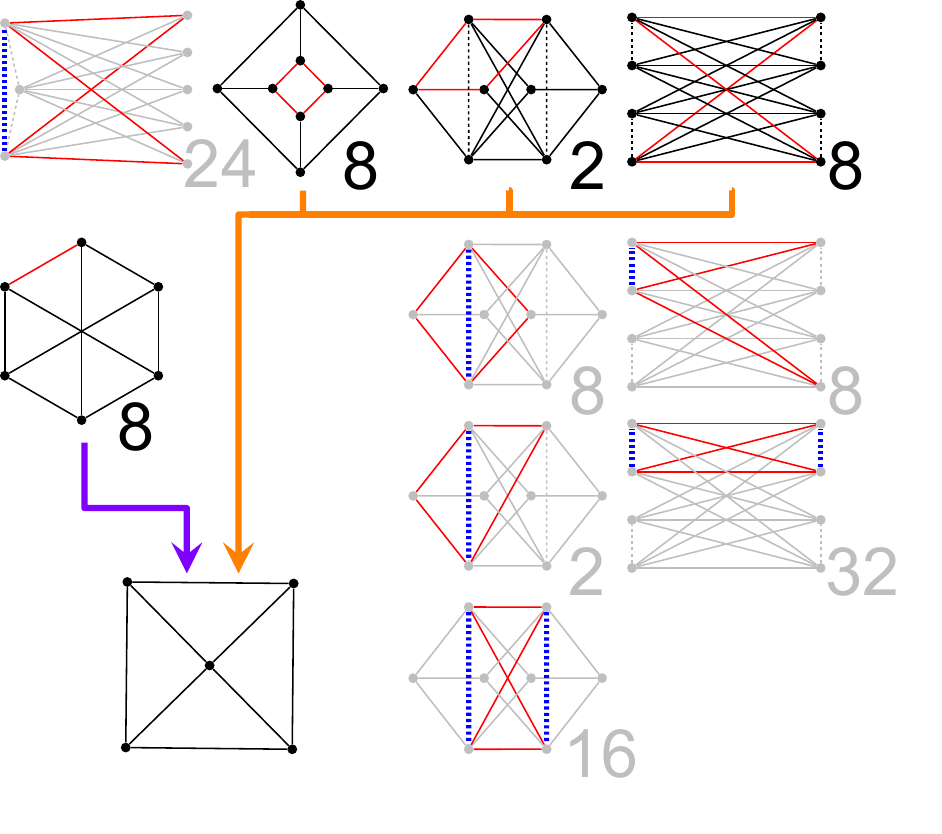}}
    \subcaptionbox{Double-quadrangle rule:\\$\beta(\frac{c_3}2+\frac{c_4}4)=\frac{b_1}4$}{\includegraphics[width=0.48\linewidth]{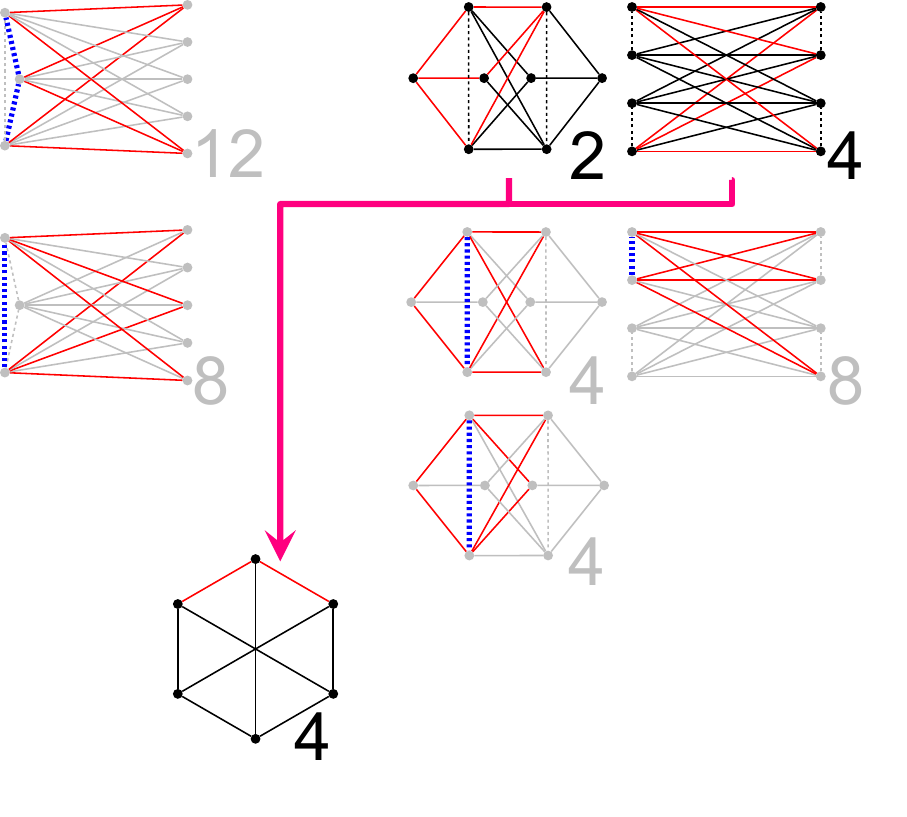}}
    \caption{The (double-)quadrangle rule relating $F_8$ and $F_6$, with the symmetry factors shown beside the highlighted $f$-graphs. The highlighted $f$-graphs in {\color{lightgray}lightgray} are mapped to 0 by ${\color{qcolor}\mathcal Q}$ or ${\color{dcolor}\mathcal D}$ due to the numerators colored in {\color{blue}blue}; hence they do not contribute to the constraining equations. Note that there is no way of highlighting a double-quadrangle in $f_8^2{=}\includegraphics[align=c,scale=0.07]{images/bf8_2.pdf}$.}
    \label{fig:f86}
\end{figure}

Drawing analogies, it is natural to look for graphical rules relating $F_N$ and $F_{N-2}$ in ABJM since $N$ is even. Since the number of bipartite $f$-graphs is much smaller than that of planar $f$-graphs, we look for graphical rules in the former representation. Only even-length solid line cycles are allowed in bipartite $f$-graphs, so instead of triangles, we look for quadrangles, which motivates the following putative graphical rules. Given $F_N{=}\sum_ic_if_N^i$ and $F_{N-2}{=}\sum_jb_jf_{N-2}^j$ where $f_N^i,f_{N-2}^j$ are the bipartite $f$-graphs, the \textbf{quadrangle rule} reads (Fig.~\ref{fig:f86}(a))
\begin{equation}
    \alpha\sum_ic_i\sum_{\square\in f_N^i}\frac{{\color{qcolor}\mathcal Q}(f_N^{i,\square})}{|f_N^{i,\square}|}=2\sum_jb_j\sum_{/\in f_{N-2}^j}\frac{{\color{ecolor}\mathcal E}(f_{N-2}^{j,/})}{|f_{N-2}^{j,/}|}.
\end{equation}
Here, $f_N^{i,\square}$ means $f_N^i$ with a (solid line) quadrangle highlighted (i.e., the corresponding $x_{ij}^2{=}0$), and similarly for the edge-highlighted $f_{N-2}^{j,/}$; $|f_N^{i,\square}|$ and $|f_{N-2}^{j,/}|$ denote the symmetry factors of the highlighted $f$-graphs. Since the graphical rule is putative, we allow for an arbitrary overall factor $\alpha$ to be determined by consistency. ${\color{qcolor}\mathcal Q}$ shrinks the quadrangle and ${\color{ecolor}\mathcal E}$ shrinks the edge down to a point:
\begin{equation}
    {\color{qcolor}\mathcal Q:}\ \includegraphics[scale=0.4,align=c]{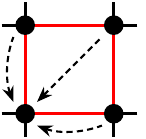}\ {\color{qcolor}\mapsto}\ \includegraphics[scale=0.4,align=c]{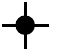},\quad{\color{ecolor}\mathcal E:}\ \includegraphics[scale=0.4,align=c]{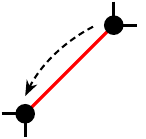}\ {\color{ecolor}\mapsto}\ \includegraphics[scale=0.4,align=c]{images/QEafter.pdf}.
\end{equation}
Comparing Fig.~\ref{fig:f86}(a) with the correct results~\eqref{eq:F8b} and~\eqref{eq:M60}, we see that choosing $\alpha{=}1$ is sufficient to fix the relative coefficient $c'{=}{-}2c$ in~\eqref{eq:F8boots}. The \textbf{double-quadrangle rule} reads (Fig.~\ref{fig:f86}(b))
\begin{equation}
    \beta\sum_ic_i\sum_{\dq\in f_N^i}\frac{{\color{dcolor}\mathcal D}(f_N^{i,\dq})}{|f_N^{i,\dq}|}=\sum_jb_j\sum_{<\in f_{N-2}^j}\frac{f_{N-2}^{j,<}}{|f_{N-2}^{j,<}|},
\end{equation}
where $\beta$ is another overall factor and ${\color{dcolor}\mathcal D}$ pinches the highlighted double-quadrangle to a highlighted cusp:
\begin{equation}
    {\color{dcolor}\mathcal D:}\ \includegraphics[scale=0.4,align=c]{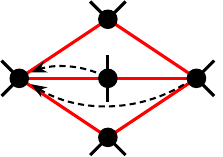}\ {\color{dcolor}\mapsto}\ \includegraphics[scale=0.4,align=c]{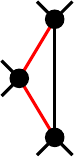}.
\end{equation}
Comparing Fig.~\ref{fig:f86}(b) with~\eqref{eq:F8b} and~\eqref{eq:M60}, we see that it is consistent to choose $\beta{=}\alpha/2{=}1/2$. In fact, a weaker version of the double-quadrangle rule exactly corresponds to the double-soft limit~\cite{Chin:2015qza} of the squared amplitude~\footnote{This is similar to SYM, where the double-triangle rule has a weaker version corresponding to the soft limit of the squared amplitude.}.

With these putative graphical rules, we bootstrap $F_{10}{\in}\mathcal B{\cap}\mathcal P$. There are 19741 planar and 120 bipartite $f$-graphs in total, and we numerically find $3{=}\dim\mathcal B{\cap}\mathcal P$ linear relations relating them. Finally, the quadrangle and double-quadrangle rules uniquely fix the answer in the bipartite representation modulo Gram identities, which we present in an ancillary file.

It is extremely nontrivial that the putative graphical rules have a unique consistent solution $F_{10}$. We have numerically checked that $M_{10}^{(0)}{=}\lim\sigma_{10}F_{10}$ matches the correct result obtained from the Grassmannian formula~\cite{Lee:2010du} (see the Supplemental Material for details). We have also extracted $A_4^{(6)}$ by subtracting off $A_4^{(\ell)}A_4^{(6-\ell)}$ ($1{\leq}\ell{\leq}5$)~\cite{He:2022cup} from $M_4^{(6)}$ and verified that $A_4^{(6)}$ has correct unitarity cuts~\cite{He:2023exb}. We leave the check against $M_8^{(2)}$~\cite{He:2022lfz} as well as further studies of these new data at $L{=}4,6$ to future works.

\section{Conclusion}

In this Letter, we observed that the (planar integrands of) ABJM squared amplitudes, $M_n^{(L)}$, can be unified in an $S_N$-permutation-invariant generating function $F_N$ ($N{:=}n{+}L$), which satisfies planarity as well as a beautiful bipartite property. These two basic properties combined already strongly constrain $F_N$; paralleling the amplitude/correlator duality in SYM, we proposed putative graphical rules which we used to bootstrap it up to $F_{10}$ providing valuable perturbative data for squared amplitudes. Our results strongly suggests the existence of a ``bipartite correlator'' dual to amplitudes in ABJM, which should exhibit universal behaviors under OPE and cusp limits similar to the half-BPS correlators in SYM.

Conceptually, it is extremely important to search for the exact physical observable behind the generating function $F_N$. The permutation symmetry unifying loops and legs suggests it should be related to ``Lagrangian insertion'' in ABJM. However, since $k_{\rm CS}$ is quantized, there is no exactly marginal scalar operators respecting $\mathcal N{=}6$ superconformal symmetry~\cite{Cordova:2016emh}, and the SYM mechanism $\int{\rm d}y\langle{\cdots}\mathcal L(y)\rangle{=}a\partial_a\langle{\cdots}\rangle$ likely has to be modified. On the other hand, the manifest $SU(3)$ R-symmetry in $\mathcal A_n$ suggests that $F_N$ might be related to correlators in mass-deformed ABJM~\cite{Benna:2008zy} which breaks superconformal symmetry to $\mathcal N{=}2$ but exhibits a global $SU(3)$ symmetry. Identifying the correlator behind $F_N$ is necessary if we wish to justify the graphical rules and pushing the bootstrap to higher $N$ like in SYM. It will also shed light on properties of ABJM and its relation with SYM.

One immediate question is how to efficiently extract the (un-squared) amplitudes from $M_n^{(L)}$ for $n\geq 6$ along the lines of~\cite{Heslop:2018zut}. Geometrically, it is interesting to look for the three-dimensional ``squared amplituhedron'' and ``correlahedron''~\cite{Eden:2017fow,Dian:2021idl,He:2024xed,He:2025rza} in ABJM. From our results, one can also extract the collinear splitting functions in ABJM (analogous to~\cite{He:2024hbb} for SYM), which can be exploited for other important quantities such as energy correlators. 

Last but not least, the weight-3 $f$-graphs are also mathematically interesting since they generate a vast collection of interesting three-dimensional DCI integrals, {\it e.g.} some of these are known to evaluate to elliptic functions (see~\cite{He:2023qld}). As graphical functions~\cite{Borinsky:2021gkd}, it is also worth studying their periods
\begin{equation}
    \mathcal P_{f_N^i}\propto\int\frac{{\rm d}^3x_1\cdots{\rm d}^3x_N}{SO(5)}f_N^i,
\end{equation}
which would provide valuable data in number theory~\cite{Broadhurst:1995km,Schnetz:2008mp,Brown:2009ta,Schnetz:2013hqa,Panzer:2016snt,He:2025vqt,He:2025lzd} and regarding the corresponding ``integrated correlators''(see ~\cite{Wen:2022oky,Brown:2023zbr,Zhang:2024ypu} for SYM case) in ABJM theory.

\begin{CJK*}{UTF8}{gbsn}
\section*{Acknowledgments}
It is our pleasure to thank Jacob Bourjaily, Clay C\'ordova, Xuhang Jiang (姜旭航), Chia-Kai Kuo ({\CJKfamily{bsmi}郭家愷}), Xiang Li (李想), Authur Lipstein, and Jiahao Liu (刘家昊) for discussions. SH has been supported by the National Natural Science Foundation of China under Grant No. 12225510, 12047503, 12247103, and by the New Cornerstone Science Foundation through the XPLORER PRIZE. The work of CS is supported by the China Postdoctoral Science Foundation under Grant No. 2022TQ0346, and the National Natural Science Foundation of China under Grant No. 12347146.
\end{CJK*}

\bibliographystyle{physics}
\bibliography{reference.bib}

\appendix
\widetext
\begin{center}
    \textbf{\large Supplemental Material}
\end{center}

\section{Summary of graphical rules in SYM}

We summarize the constraints on $F_N$ and the corresponding graphical rules in SYM. This will make it clear why the putative ABJM graphical rules put forth in the main text are of the given form.

\subsection{The triangle rule}

The triangle rule arises from the physical OPE limit~\cite{Eden:2012tu} of the four-point half-BPS correlator, which at the integrand level translates to the following condition (see~\cite{He:2024cej} for a detailed derivation):
\begin{equation}
    \lim_{x_2,x_N\to x_1}x_{12}^2x_{1N}^2x_{2N}^2F_N=6\lim_{x_2\to x_1}x_{12}^2F_{N-1}.
\end{equation}
Suppose $F_N=\sum_ic_if_N^i$ and $F_{N-1}=\sum_jb_jf_{N_1}^j$ are the corresponding $f$-graph representations. In order to survive the limit on the LHS, $f_N^i$ must contain a triangle subgraph corresponding to the denominator $\frac1{x_{12}^2x_{1N}^2x_{2N}^2}$, which is shrunk to a point. In order to survive the limit on the RHS, $f_{N-1}^j$ must contain an edge corresponding to the denominator $\frac1{x_{12}^2}$, which is shrunk to a point. Correctly accounting for the symmetry factors, we obtain
\begin{equation}
    \sum_ic_i\sum_{\triangle\in f_N^i}\frac{{\color{qcolor}\mathcal Q}(f_N^{i,\triangle})}{|f_N^{i,\triangle}|}=2\sum_jb_j\sum_{/\in f_{N-1}^j}\frac{{\color{ecolor}\mathcal E}(f_{N-1}^{j,/})}{|f_{N-1}^{j,/}|}.
\end{equation}
Here, $f_N^{i,\triangle}$ denotes $f_N^i$ with the triangle subgraph highlighted (where $\triangle$ ranges through all \emph{inequivalent} triangle subgraphs), and $f_{N-1}^{j,/}$ denotes $f_{N-1}^j$ with the edge highlighted (where $/$ ranges through all \emph{inequivalent} edges). ${\color{qcolor}\mathcal Q}$ shrinks the triangle and ${\color{ecolor}\mathcal E}$ shrinks the edge down to a point:
\begin{equation}
    {\color{qcolor}\mathcal Q:}\ \includegraphics[scale=0.4,align=c]{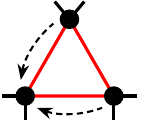}\ {\color{qcolor}\mapsto}\ \includegraphics[scale=0.4,align=c]{images/QEafter.pdf},\quad{\color{ecolor}\mathcal E:}\ \includegraphics[scale=0.4,align=c]{images/Ebefore.pdf}\ {\color{ecolor}\mapsto}\ \includegraphics[scale=0.4,align=c]{images/QEafter.pdf}.
\end{equation}

\subsection{The double-triangle rule}

\newcommand{\dt}{\triangleleft\!\triangleright}

The double-triangle rule arises from the physical cusp limit~\cite{He:2024cej}, a generalization of the null-polygonal lightlike limit~\cite{Alday:2010zy} of the four-point half-BPS correlator, which at the integrand level translates to the following condition (see~\cite{He:2024cej} for a detailed derivation):
\begin{equation}
    \lim_{\substack{x_{12}^2,x_{23}^2\to 0\\x_N\to x_2}}\frac{x_{12}^2x_{23}^2x_{1N}^2x_{2N}^2x_{3N}^2}{x_{13}^2}F_N=2\lim_{x_{12}^2,x_{23}^2\to 0}x_{12}^2x_{23}^2F_{N-1}.
\end{equation}
Suppose $F_N=\sum_ic_if_N^i$ and $F_{N-1}=\sum_jb_jf_{N_1}^j$ are the corresponding $f$-graph representations. In order to survive the limit on the LHS, $f_N^i$ must contain a double-triangle subgraph corresponding to the denominator $\frac1{x_{12}^2x_{23}^2x_{1N}^2x_{2N}^2x_{3N}^2}$, which is pinched to a cusp. In order to survive the limit on the RHS, $f_{N-1}^j$ must contain a cusp corresponding to the denominator $\frac1{x_{12}^2x_{23}^2}$. Correctly accounting for the symmetry factors, we obtain
\begin{equation}
    \sum_ic_i\sum_{\dt\in f_N^i}\frac{{\color{dcolor}\mathcal D}(f_N^{i,\dt})}{|f_N^{i,\dt}|}=\sum_jb_j\sum_{<\in f_{N-1}^j}\frac{f_{N-1}^{j,<}}{|f_{N-1}^{j,<}|}.
\end{equation}
Here, $f_N^{i,\dt}$ denotes $f_N^i$ with the double-triangle subgraph highlighted (where $\dt$ ranges through all \emph{inequivalent} double-triangle subgraphs), and $f_{N-1}^{j,<}$ denotes $f_{N-1}^j$ with the cusp highlighted (where $<$ ranges through all \emph{inequivalent} cusps). ${\color{dcolor}\mathcal D}$ pinches the highlighted double-triangle to a highlighted cusp:
\begin{equation}
    {\color{dcolor}\mathcal D:}\ \includegraphics[scale=0.4,align=c]{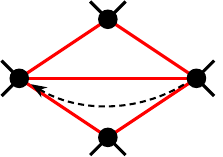}\ {\color{dcolor}\mapsto}\ \includegraphics[scale=0.4,align=c]{images/Dafter.pdf}.
\end{equation}

\section{Numerical computation of squared tree amplitudes from the Grassmannian formula}

Package the kinematic data $\lambda_i^\alpha$ (resp., $\eta_i^I$) into a $2\times n$ (resp., $3\times n$) matrix, or a 2-plane (resp. 3-plane) in $\mathbb C^n$. For later convenience, we will define a Wick-rotated kinematics as:
\begin{equation}
    \boldsymbol\lambda_i^\alpha|\boldsymbol\eta_i^I:=\lambda_i^\alpha|\eta_i^I\times\begin{cases}
        1,&i\text{ odd};\\{\rm i},&i\text{ even}.
    \end{cases}
\end{equation}
Since $n=2k$ is even, we can introduce a $(k,k)$-signature metric $g^{ij}={\rm diag}(+1,-1,\cdots,+1,-1)$. Following the conventions in~\cite{Huang:2014xza}, the Grassmannian formula for ABJM tree superamplitudes (first proposed in~\cite{Lee:2010du}) reads:
\begin{equation}\label{eq:grassmannian}
\begin{aligned}
    \mathcal A_{n=2k}=\sum_\sigma\Big[&\oint_{\sigma_+}\frac{{\rm d}^{k\times2k}C_+}{GL(k)}\frac{\delta^{\frac{k(k+1)}2}(C_+\cdot C_+^T)}{(1\cdots k)(2\cdots(k{-}1))\cdots(k\cdots(2k{-}1))}\delta^{k\times2}(C_+\cdot\boldsymbol\lambda^T)\delta^{k\times3}(C_+\cdot\boldsymbol\eta^T)\\
    +\,&\oint_{\sigma_-}\frac{{\rm d}^{k\times2k}C_-}{GL(k)}\frac{\delta^{\frac{k(k+1)}2}(C_-\cdot C_-^T)}{(1\cdots k)(2\cdots(k{-}1))\cdots(k\cdots(2k{-}1))}\delta^{k\times2}(C_-\cdot\boldsymbol\lambda^T)\delta^{k\times3}(C_-\cdot\boldsymbol\eta^T)\Big].
\end{aligned}
\end{equation}
where $C_\pm$ is the matrix representative of an element in the positive/negative orthogonal Grassmannian $OG_\pm(k,2k)$, whose $GL(k)$ redundancy can be gauge-fixed as:
\begin{equation}
    C_{\pm\alpha i}=\begin{cases}
        1,&\text{$i$ odd \& $i=2\alpha-1$};\\0,&\text{$i$ odd \& $i\neq2\alpha-1$};\\c_{\pm\alpha\beta},&\text{$i$ even \& $i=2\beta$},
    \end{cases}
\end{equation}
where $\det[c_{\pm\alpha\beta}]_{\alpha,\beta=1}^k=\pm1$. Without loss of generality, we choose $c_{-\alpha\beta}$ as $c_{+\alpha\beta}$ with the last column sign-flipped:
\begin{equation}
    c_-=c_+\left(\begin{matrix}1\\&\ddots\\&&1\\&&&-1\end{matrix}\right).
\end{equation}
Equivalently, we can think of $C_\pm$ as a null $k$-plane in $\mathbb C^{k,k}$. The dot product denotes contractions with the metric $g^{ij}$.

At tree level, the integration is performed over certain BCFW cells labeled by perfect matchings $\sigma$ of $\{1,\cdots,n\}$ with codimension $\frac12(k-2)(k-3)$, or dimension $2k-3$. For $n\leq10$ (which is all we need), these correspond to the vanishing of $\frac12(k-2)(k-3)$ minors in the denominator of~\eqref{eq:grassmannian}. Following the detailed prescription in~\cite{Huang:2014xza}, we can easily obtain a parametrization of $c_+$ (and hence $c_-$) in terms of $2k-3$ angle variables $\theta_s$ from any medial graph representation of $\sigma$ such that
\begin{equation}\label{eq:para}
    \oint_{\sigma_\pm}\frac{{\rm d}^{k\times 2k}C_\pm}{GL(k)}\frac{\delta^{\frac{k(k+1)}2}(C_\pm\cdot C_\pm^T)}{(1\cdots k)(2\cdots(k{-}1))\cdots(k\cdots(2k{-}1))}=\mathcal J_\pm\prod_{s=1}^{2k-3}\,{\rm d}\log\tan\theta_s=\mathcal J_\pm\prod_{s=1}^{2k-3}\frac{1+\tau_s^2}{\tau_s(1-\tau_s^2)}{\rm d}\tau_s,\quad\tau_s:=\tan\frac{\theta_s}2.
\end{equation}
Here, the absolute value of $\mathcal J_\pm$ are both given by
\begin{equation}
    |\mathcal J_\pm|=1+\mathcal J_1+\cdots=\text{eq.(2.25) of~\cite{Huang:2014xza},}
\end{equation}
but the signs of $\mathcal J_\pm$ have not been well-documented in the literature. However, it is very important that the signs are chosen consistently across all $\sigma_\pm$ in~\eqref{eq:grassmannian}. In practice, we compute $\mathcal J_\pm\prod\frac{1+\tau_s^2}{\tau_s(1-\tau_s^2)}$ as follows:
\begin{enumerate}
    \item Split the $k^2$ variables $c_{\pm\alpha\beta}$ into two subsets: $\{c'\}$ with $\frac{k(k+1)}2+\frac{(k-2)(k-3)}2$ elements, and $\{c''\}$ with $(2k-3)$ elements. This split has to be the same for all $\sigma_\pm$.
    \item Compute the Jacobian determinant $\mathcal J'=|\frac{\partial(\text{constraints})}{\partial c'}|$ relating the constraints (namely, the $\frac{k(k+1)}2$ delta function arguments and $\frac12(k-2)(k-3)$ vanishing minors) to the variables $\{c'\}$.
    \item Compute the Jacobian determinant $\mathcal J''=|\frac{\partial c''}{\partial\tau_s}|$ relating the variables $\{c''\}$ and the parametrization $\{\tau_s\}$.
    \item The total Jacobian is given by $\mathcal J_\pm\prod\frac{1+\tau_s^2}{\tau_s(1-\tau_s^2)}=\mathcal J''/\mathcal J'$.
\end{enumerate}

The next step is to localize the $(2k-3)$-form~\eqref{eq:para} using $\delta^{k\times 2}(C_\pm\cdot\boldsymbol\lambda^T)$ in~\eqref{eq:grassmannian} to obtain a distribution supported by momentum conservation $\delta^3(\boldsymbol\lambda\cdot\boldsymbol\lambda^T)$. The solutions $\tau_s^*$ can be easily obtained numerically using \texttt{NSolve[]} in Mathematica to very high precision (thousands of digits), but to compute the Jacobian is highly nontrivial which has not been well-documented in the literature either. We overcome this difficulty by viewing the superamplitude~\eqref{eq:grassmannian} as a differential form $\mathcal A_n\frac{{\rm d}^{2\times n}\boldsymbol\lambda}{SL(2)}$ in kinematic space~\cite{Arkani-Hamed:2012zlh}. We can fix the $SL(2)$-redundancy by fixing $\boldsymbol\lambda$ to be
\begin{equation}
    \hat{\boldsymbol\lambda}=\left(\begin{matrix}\ell&p&0&\hat{\boldsymbol\lambda}_{i>3}^1\\0&q&\ell&\hat{\boldsymbol\lambda}_{i>3}^2\\\end{matrix}\right).
\end{equation}
With this choice, formally we can write
\begin{equation}\label{eq:k0}
    \mathcal A_n\frac{{\rm d}^{2\times n}\boldsymbol\lambda}{SL(2)}=\frac{\mathcal A_n}{\delta^3(\boldsymbol\lambda\cdot\boldsymbol\lambda^T)}\delta^3(\boldsymbol\lambda\cdot\boldsymbol\lambda^T)\frac{{\rm d}^{2\times n}\boldsymbol\lambda}{SL(2)}=\frac{\mathcal A_n}{\delta^3(\hat{\boldsymbol\lambda}\cdot\hat{\boldsymbol\lambda}^T)}\frac1{\mathcal K^\circ}{\rm d}^{2\times(n-3)}\hat{\boldsymbol\lambda},\quad\mathcal K^\circ=\left|\frac{\partial^3(\hat{\boldsymbol\lambda}\cdot\hat{\boldsymbol\lambda}^T)}{\partial(\ell,p,q)}\right|=4\ell(p^2+q^2).
\end{equation}
On the other hand, due to $GL(k)$ invariance,
\begin{equation}\label{eq:rho}
    \frac{{\rm d}^{2\times 2k}\boldsymbol\lambda}{SL(2)}\frac{{\rm d}^{k\times 2k}C}{GL(k)}\delta^{k\times 2}(C\cdot\boldsymbol\lambda^T)=\frac{{\rm d}^{2\times 2k}\boldsymbol\lambda\,{\rm d}^{2\times k}\rho\,{\rm d}^{k\times 2k}C}{SL(2)\times GL(k)}\delta^{2\times2k}(\rho C-\boldsymbol\lambda),
\end{equation}
where $\rho_{2\times k}$ intertwines the $GL(k)$ with the kinematic $SL(2)$. Now, we have
\begin{equation}\label{eq:k}
    {\rm d}^{2k-3}\tau\,\delta^{k\times 2}(C\cdot\boldsymbol\lambda^T)\frac{{\rm d}^{2\times n}\boldsymbol\lambda}{SL(2)}={\rm d}^{2k-3}\tau\int{\rm d}^{2\times k}\rho\,\delta^{2\times2k}(\rho C-\boldsymbol\lambda)\frac{{\rm d}^{2\times n}\boldsymbol\lambda}{SL(2)}=\frac1{\mathcal K}{\rm d}^{2\times(n-3)}\hat{\boldsymbol\lambda},\quad\mathcal K=\left|\frac{\partial^{2\times 2k}(\rho C-\hat{\boldsymbol\lambda})}{\partial(\rho_{2\times k},\tau_s,\ell,p,q)}\right|.
\end{equation}
The delta function fixes $\ell=1$ and
\begin{equation}
    \rho=\hat{\boldsymbol\lambda}_{\text{odd columns}}=\left(\begin{matrix}1&0&\ast&\cdots\\0&1&\ast&\cdots\end{matrix}\right).
\end{equation}
We see that $\rho C$ just adds other rows to the first two rows, which implies
\begin{equation}
    \delta^{k\times3}(C\cdot\boldsymbol\eta^T)=\delta^{2\times3}(\rho C\cdot\boldsymbol\eta^T)\delta^{(k-2)\times3}(C_{3;;}\cdot\boldsymbol\eta^T)\xlongequal{\rho C=\boldsymbol\lambda}\delta^{2\times3}(\boldsymbol\lambda\cdot\boldsymbol\eta^T)\delta^{(k-2)\times3}(C_{3;;}\cdot\boldsymbol\eta^T),
\end{equation}
where $C_{3;;}$ denotes the matrix $C$ with the first two rows deleted. Comparing~\eqref{eq:k0} and~\eqref{eq:k}, we can now strip off $\delta^3(\boldsymbol\lambda\cdot\boldsymbol\lambda^T)\delta^{2\times3}(\boldsymbol\lambda\cdot\boldsymbol\eta^T)$ with the correct Jacobian:
\begin{equation}
    \mathcal A_n^{(0)}=\delta^3(\boldsymbol\lambda\cdot\boldsymbol\lambda^T)\delta^{2\times3}(\boldsymbol\lambda\cdot\boldsymbol\eta^T)\underbrace{\sum_{\sigma_\pm}\sum_{\text{solutions }\tau^*}\frac{\mathcal J''(\tau^*)}{\mathcal J'(\tau^*)}\frac{\mathcal K(\tau^*)}{\mathcal K^\circ(\tau^*)}\delta^{(k-2)\times3}(C_{3;;}(\tau^*)\cdot\boldsymbol\eta^T)}_{A_n^{(0)}}.
\end{equation}

To compute $M_n^{(0)}$, we expand $A_n^{(0)}$ into $\eta$-monomials and square the coefficients as described in the main text. This is easily achieved numerically, since the coefficient of an $\eta$-monomial in $\delta^{(k-2)\times3}(C_{3;;}\cdot\boldsymbol\eta^T)$ is nothing but a minor of $C_{3;;}$ (with an additional factor of ${\rm i}$ for every $\eta_i^I$ with even $i$, which arises from the difference between $\boldsymbol\eta$ and $\eta$).

\end{document}